# Slime Mould Logic Gates Based on Frequency Changes of Electrical Potential Oscillation

James G.H. Whiting [1], Ben P.J. de Lacy Costello [1,2], Andrew Adamatzky [1].

[1] *Unconventional Computing Centre, University of the West of England, Bristol, UK.*
[2] *Institute of Biosensing Technology, University of the West of England, Bristol, UK.*

**Abstract**

*Physarum polycephalum* is a large single amoeba cell, which in its plasmodial phase, forages and connects nearby food sources with protoplasmic tubes. The organism forages for food by growing these tubes towards detected food stuffs, this foraging behaviour is governed by simple rules of photoavoidance and chemotaxis. The electrical activity of the tubes oscillates, creating a peristaltic like action within the tubes, forcing cytoplasm along the lumen; the frequency of this oscillation controls the speed and direction of growth. External stimuli such as light and food cause changes in the oscillation frequency. We demonstrate that using these stimuli as logical inputs we can approximate logic gates using these tubes and derive combinational logic circuits by cascading the gates, with software analysis providing the output of each gate and determining the input of the following gate. Basic gates OR, AND and NOT were correct 90%, 77.8% and 91.7% of the time respectively. Derived logic circuits XOR, Half Adder and Full Adder were 70.8%, 65% and 58.8% accurate respectively. Accuracy of the combinational logic decreases as the number of gates is increased, however they are at least as accurate as previous logic approximations using spatial growth of *Physarum polycephalum* and up to 30 times as fast at computing the logical output. The results shown here demonstrate a significant advancement in organism-based computing, providing a solid basis for hybrid computers of the future.

Introduction

*Physarum polycephalum* is a single celled organism visible by the unaided eye [1]. It can span several of centimetres in warm, dark and humid conditions, and has been discovered engulfing decomposing trees. When inoculated in the environment with distributed sources of nutrients, the slime mould spans the nutrients using a network of protoplasmic tubes. The topology of the network is controlled by gradients of chemical, optical and thermal attractants and repellents [2], [3]. The flow of cytoplasm within these protolasmic networks is governed by shuttle streaming [4], [5] which controls locomotion and growth [6], [7], [8]; rhythmic contractions controlled by $Ca^{2+}$ oscillations [9], [10], [11] cause the actin myosin filaments to initiate peristaltic action of the cell membrane.

The oscillatory period of shuttle streaming is between 60 and 200 seconds [12], [13], [14]; this frequency is increased and decreased by attractant and repellent stimuli respectively [3], [15], [16][17]. Chemotaxis is exhibited by *P. polycephalum* with food sources providing reliable attractant sources [18], [19], [20], [21], [22], and heat [3], [13], [19], [23] also having attractant properties; *P. polycephalum* avoids light sources [24], [25], [26], [27] and certain chemicals [15], [16], [28], [29].

*P. polycephalum* responds to certain stimuli, this response can be measured either by observing growth or by electronic measurement of the surface electrical potential [12], [16], [30], [31]. The recordable electrical response of *P. polycephalum* to these stimuli is repeatable; thus several biosensors using *P. polycephalum* have been developed [16], [32], [33], [34].

The main goal of *P. polycephalum* computing is to design a general purpose processing device from the plasmodial phase of slime mould [35], [36]. Thus implementation of logical gates is the most important task of *P. polycephalum* computing. So far two types of *P. polycephalum* logic gates have been implemented with living *P. polycephalum* in laboratory experiments [37], [38] and simulations [39].

Tsuda, Aono and Gunji [38] developed the first ever *P. polycephalum* logic gates based on the slime mould's propagation along agar channels in specially laid out geometry. Their gates use the presence and absence of Physarum in a given loci of space for logic values 1 and 0 respectively. The *P. polycephalum* gates use both the foraging nature and plasmodial fusion avoidance properties to compute the logic calculation, with the agar geometry determining the type of gate. The accuracy of Tsuda et al. gates AND, NOT and OR are 69%, 83% and 100% respectively. One factor to consider however, is the computation time, Tsuda et al. documented that time to completion ranged from 11 to 18 hours per gate; the limitation is a direct result of the speed of growth.

Adamatzky [37] used a collision-based computing approach [40] to implement *P. polycephalum* gates. On a non-nutrient substrate the plasmodium propagates as a traveling localization, in the form of a compact wave-fragment of protoplasm; this Physarum-localization travels in its originally predetermined direction for a substantial period of time, even when no gradient of chemo-attractants is present. Adamatzky [37] utilized this property of *P. polycephalum* active growing zones to design two-input two-output Boolean logic gates <X> --> <x AND y, x OR y> and <x,y> -> <x, NOT x AND y> verifying the designs in laboratory experiments and computer simulations, cascading the logical gates into a one-bit half-adder and simulating its functionality. Accuracy of experimental laboratory prototypes of gate <x,y> --> <x AND y, x OR y> was over 69% and of gate <x,y> -> <x, NOT x AND y> over 59%. The accuracy was comparable to the gates in the experiments of Tsuda et al. [38]. Jones and Adamatzky [39] designed yet another modification of collision-based, or ballistic logical gates and produced a 1-bit half adder using experimentally derived computational modelling.

With the right substrate geometry and attractant-repellent manipulation, *P. polycephalum* growth can calculate the output of basic and derived logic gates, with an acceptable degree of accuracy. The limitation with all these schemes is, as mentioned above, speed of growth and hence calculation time. In a previous paper [17], we presented a scheme of Physarum logical gates which used frequency change of shuttle streaming in response to stimuli to simulate the outputs of the basic logic gates AND, OR and NOT  (Table 2); while these are not strictly logic gates as they have non-equivalent input and outputs, they show a significant improvement of computation time using *P. polycephalum* as a computation medium, with calculation time reduced to 20-30 minutes. Unconventional computing with *P. polycephalum* is significantly advanced with this scheme, and this paper presents expansion of this scheme into one that defines derived and combinational logic using stimuli frequency change in *P. polycephalum* using cascaded *P. polycephalum* logic gates whose output is evaluated by software analysis which determines the input of the following gate.

**Method**

*Physarum polycephalum* culture

*Physarum polycephalum* plasmodium was grown on non-nutrient 2% agar gel in 9 centimetre diameter Petri dishes (Fisher Scientific, UK) fed daily with Organic rolled oats (Waitrose, UK). The plasmodium was moved to new agar-filled Petri dishes every week to limit unwanted microbial growth.

*P. polycephalum* response to stimuli

Data on *P. polycephalum's* response to stimuli was collected for previous research and is described in full in [17]. The data collected is processed and presented here (table 1) in order to derive additional *P. polycephalum* based Gates. Figure 1 shows the experimental set up in order to produce and measure a single protoplasmic tube. 1ml of non-nutrient Agar is placed on each of the aluminium electrodes (Farnell, UK) in a customised 9cm Petri dish (Fisher Scientific, UK) to form a cell interface. A *Physarum polycephalum* inoculated oat flake from culture is placed on 1 agar hemisphere while a bare oat flake is placed on the remaining agar hemisphere. The agar acts as a growth medium for the organism on the electrode. After a minimum of 5 hours and maximum of , a single protoplasmic tube grows between the two electrodes, allowing recording of the surface potential of the tube. Electrical measurement of the protoplasmic tube were performed by connecting the aluminium electrodes to a PicoLog ADC-24 high resolution analogue-to-digital data logger (Pico Technology, UK) connected via USB to a laptop installed with PicoLog Recorder software for data capture. The PicoLog ADC-24 recorded ±39 millivolts at 1 hertz for the duration of the experiment, with a 24 bit resolution; the originally inoculated agar hemisphere was connected to ground, while the newly connected agar hemisphere was connected to an analogue recording channel.

Stimulation of the organism was performed by adding an oat flake on the recording electrode or by heating the recording electrode to 10°C above room temperature using a 1.4W Peltier element (RS Components, UK) placed underneath the Petri dish at the site of the recording electrode. Simultaneous heating and oat flake addition was also performed. The 10 minute period before stimulation was used as the baseline frequency measurement ($f_{pre}$) and the 10 minutes after stimulation had started was the frequency change ($f_{post}$); frequency change($\Delta f$) was calculated as demonstrated by equation (1) and expressed as a percentage.

$$\Delta Freq = {f_{pre}}/{f_{post}} \tag{1}$$

Logic gate approximation as described in [17] uses frequency change of shuttle streaming to determine a logic True or False output. A logic gate is a single protoplasmic tube of *P. polycephalum* (figure 1), which is stimulated with combinations of inputs, light, oat flakes or heat. The frequency change is measured by custom Matlab software which performs a frequency analysis before and after the stimuli, determining the logical output; the type of gate used determines the thresholds for logic 1 or 0, as shown in tables 2 and 3. Using the principals of combinational logic, this paper uses the basic gates to produce derived gates NOR, NAND, XOR and XNOR, as well as more complex combinational logic circuits. Combinational logic circuits are cascaded manually, with the software output detailing the input for the following gate which is performed manually; in the future it is envisioned that this process can be totally automated. For the logic gate inputs A and B, the stimuli Heat and Oat Flake was applied respectively. While A and B are both false (or logic 0), neither the heat nor the oat flake is applied, while A and B are both true (or logic 1) both stimuli are simultaneously applied. The frequency change for each gate is shown in table 1 and the classification for the previously produced simple logic gates are detailed in table 2.

Table 1. Frequency change in response to stimuli and equivalent logic input

| Stimuli Type | Logic input (A,B) | Median Frequency Change | Standard Deviation |
|---|---|---|---|
| No stimulus | 0,0 | 2.1% | 6.9 |
| Oat Flake | 0,1 | 12.2% | 12.6 |
| Heat | 1,0 | 19.8% | 8.8 |
| Heat and Oat Flake | 1,1 | 33.2% | 9.6 |

Table 2. Type of logic gate implemented is dependent on thresholds of frequency change, $\Delta f$:

OR gate.    $\Delta f < 10\%$, Logic 0.    $\Delta f \geq 10\%$, Logic 1.

AND gate.    $\Delta f < 24\%$, Logic 0.    $\Delta f \geq 24\%$, Logic 1.

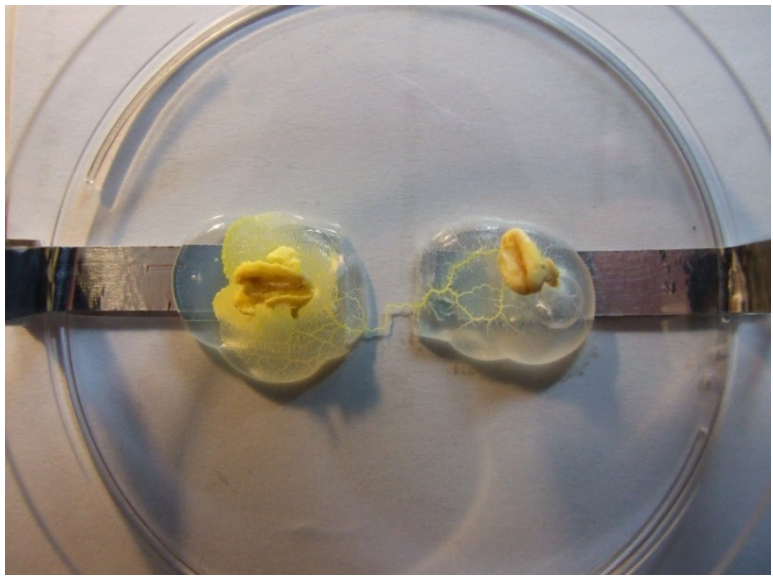

Figure 1. A single protoplasmic tube branching two agar hemispheres. Logic computation is performed by inducing frequency change of shuttle streaming along this tube; the stimuli are applied to the left hemisphere.

**Results**

*XOR* gate

The Exclusive OR (XOR) gate is commonly used in binary adders and other logic circuits; the output is high if either input is True but not both, otherwise the output is False. A frequency change system can be deduced using two thresholds, in a similar manner to that published previously [17]. Frequency change of between 4.9% and 32% (inclusive) is logical True or 1, a change of either less than 4.9% or greater than 32% is a logical False or 0 (table 3).

*NAND, NOR, XNOR* gates

While the AND, OR and XOR gates calculate specific outputs, they can be inverted by, producing NAND, NOR and XNOR gates respectively. These gates normally have NOT gates at each input or a single NOT gate at the output, so are in essence combinational logic. Hence the frequency system for each gate can be simply modified by inverting the threshold categories (table 3), for example, an OR gate is high when the frequency change is greater or equal to 10%, and low when less than 10%; a NOR gate is low when the frequency change is greater or equal to 10%, and high when less than 10%. Alternatively the inputs can be inverted, and the inputs to the OR gate are high when present, that is, when an oat flake and heat are on, whereas for the NOR gate, the inputs are high when the oat flake and heat are not present; this becomes more useful when some inputs are inverted and others aren't as in the 2-4 bit Decoder (figure 2).

Table 3. Derived logic gates and the threshold of frequency change, $\Delta f$

| Gate | | |
|---|---|---|
| NOR gate. | $\Delta f < 10\%$, Logic 1. | $\Delta f \geq 10\%$, Logic 0. |
| NAND gate. | $\Delta f < 24\%$, Logic 1. | $\Delta f \geq 24\%$, Logic 0. |
| XOR gate. | $\Delta f < 4.9\%$, Logic 0. | $\Delta f \geq 32\%$, Logic 1. |
| XNOR gate. | $\Delta f < 4.9\%$, Logic 1. | $\Delta f \geq 32\%$, Logic 0. |

Combinational Logic

With the production of NAND and NOR gates, functional completeness is achieved using *P. polycephalum* Frequency Gates (PFGs); any possible combination of logic gates may be produced using a combination of NAND (and NOR) gates.

Combinational circuits comprising several individual PFGs could be produced; the input stimuli can be automatically controlled using a microprocessor. The output of the PFG is measured using custom Matlab software consisting of data handling and a Fast Fourier Transform; this could be implemented on a microprocessor with analogue to digital converter and appropriate software to automatically determine frequency change hence logic output of one PFG and subsequently control the input of the next gate in the sequence. In this instance however, the frequency change was calculated using the Matlab software and the subsequent inputs were controlled manually, although frequency analysis software or hardware could be tasked with this process in the future.

The response for each stimuli type used for logic gate approximation was tested for normality. With the knowledge of frequency change distribution for each input derived from previous data[17], the probability and hence accuracy of the NOR, NAND, XOR and XNOR gates were calculated using the distributions and likelihood of either type I or II error for each input; this method accounted for the variation in response to the stimuli. In addition to the derived gates, accuracy of the Half Adder, Full Adder and 2-4 bit Decoder combinational logic circuits were also calculated. For the gates which had one inverted input and one normal (non-inverted) input such as in the 2-4 bit Decoder (figure 2), one stimuli/input state was inverted while the other one was not.

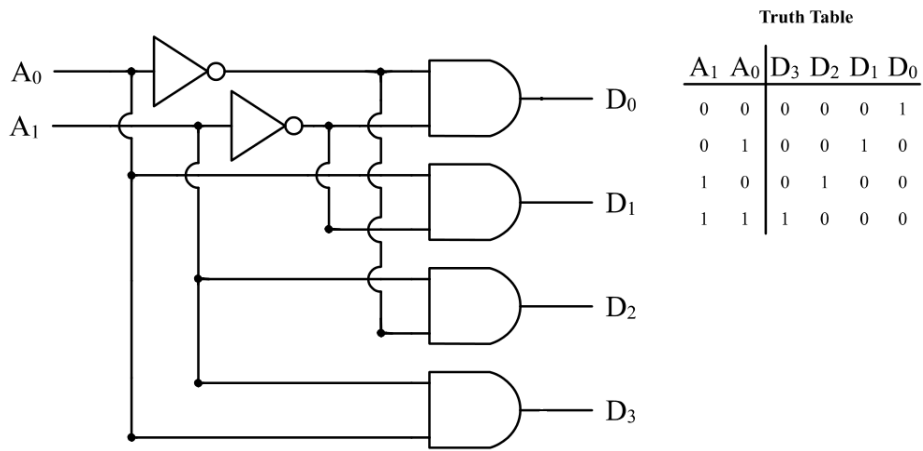

Figure 2. The 2-4 bit Decoder logic circuit with truth table.

*Inverted and combinational PFGs*

Table 4. Accuracy of combinational logic using PFGs

| Logic operation | Correct output | Number of PFGs required |
|---|---|---|
| OR/NOR | 90.0% | 1 |
| AND/NAND | 77.8% | 1 |
| NOT | 91.7% | 1 |
| XOR/XNOR | 70.80% | 1 |
| 2-4 Decoder | 57.5% | 4 |
| Half Adder | 65% | 2 |
| Full Adder | 58.80% | 5 |

The accuracy for the inverted gates is demonstrated in Table 4; with the same accuracy as the non-inverted gates presented previously [17] as the frequency change boundaries were inverted. PFGs and their inverted input has identical accuracy because only the thresholds were inverted. The accuracy of the Half Adder, Full Adder and 2-4 bit Decoder are listed in Table 4, with the least accurate being the 2-4 bit Decoder.

**Discussion**

*Physarum Frequency Gate database*

With the addition of the NAND, NOR, XOR and XNOR gates, there is now a complete database of basic and derived logic gates using frequency change. The number of inputs to these gates however is limited to 2, due to the number of tested stimuli, however both NAND and NOR gates have functional completeness, which is to say they can be combined to produce any other logical operation, including single gates with more than 2 inputs. Multi-NAND ICs are often only used in practical systems to limit the number of different chips required in a system, as multiple gates of the same type are produced on CMOS or TTL chips. The fact that the architecture of a PFG is the same regardless of gate type used, means they are effectively programmable logic gates, with the gate type being determined purely by the boundary conditions of the frequency change. The PFGs only have two inputs, however gates with more than 2 inputs can be approximated with more gates.

*PFG accuracy*

The number of PFGs used to solve a logical operation correlates with accuracy as demonstrated by figure 3; the 2-4 Decoder and Full Adder use 4 and 5 PFGs respectively and have significantly more error than the logic with 1 or 2 PFGs. The trend is not linear as the error is cumulative in a system, it is evident that logic operations with gates higher than these presented would have an accuracy no better than tossing a coin. The layout of the gates also plays a role in accuracy, as a Full Adder which has 1 more gate than a 2-4 Decoder is marginally more accurate, this is due to the series layout of the logic gates; an error produced from one gate has a chance of being coincadentally corrected by another error in the subsequent gate whereas parallel gates such as those in the 2-4 decoder are only correct if all gates produce the correct answer. In a logic system, a correct output is most important, even if the correct operation throughout the circuit is not, as in this system.

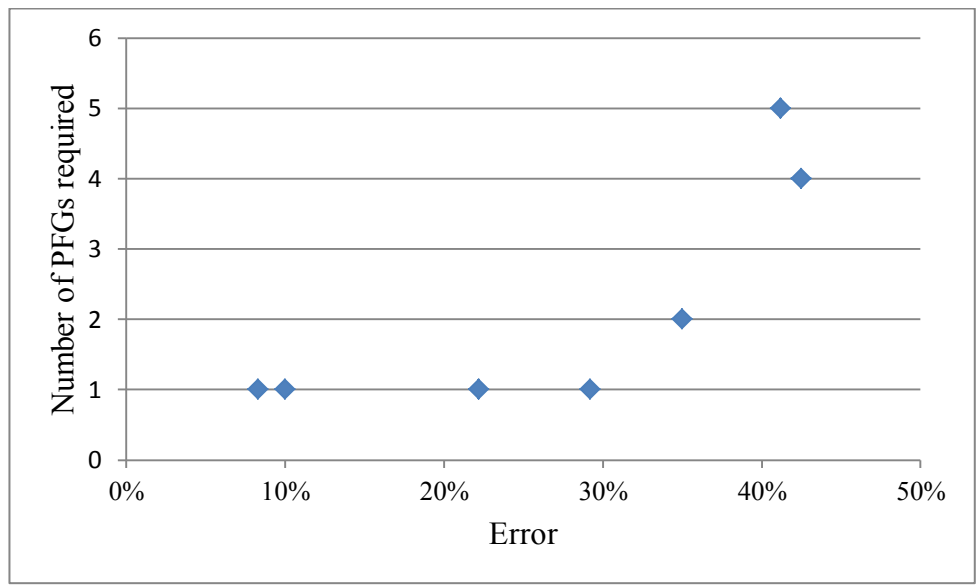

Figure 3. The correlation between the number of PFGs required to calculate a logical problem and the error rate.

The speed of processing of a PFG is 30 minutes for a single gate, decreasing the computation time by approximately 30 times compared to morphological, or growth based, gates [38], [39]. [17], the basic logic gates are as accurate as those shown in previous papers [37], [38]. The Half Adder shown in this paper is similarly accurate (65%) to that simulated by Jones and Adamatzky (63%) [39].

**Conclusion**

We have shown that frequency based Physarum Gates (PFG) can be used for approximation of Boolean logic. Functional completeness is demonstrated with the development of the NOR and NAND logic gates. NOR, NAND, XOR and XNOR derived gates have also been demonstrated; combinational logic circuits Half Adder, Full Adder and 2-4 bit Decoder have been tested with 65%, 58.8 and 57.5% accuracy respectively. Increasing the number of PFGs correlates with an increase in error of the combinational logic circuit; the error is proportional to the number of gates. A single protoplasmic tube is effectively a programmable logic gate, as the logic function is determined in the software by the type of gate used without any change to the hardware/wetware. PFGs are approximately 30 times faster than morphological, i.e. growth based, *Physarum polycephalum* logic gates.